\documentclass[pra,twoside,amsfonts,amssymb,amsmath,titlepage,floatfix,twocolumn,showpacs,superscriptaddress]{revtex4-1} 

\usepackage{amsfonts}        
\usepackage{amsmath}        
\usepackage{latexsym}
\usepackage{bm}
\usepackage{graphicx}
\usepackage{amsthm}
\usepackage{mathrsfs}
\usepackage{comment}

\usepackage{color}

\renewcommand{\phi}{\varphi}

\newcommand{\ket}[1]{|#1 \rangle}
\newcommand{\bra}[1]{\langle #1|}

\newcommand{\be}{\begin{equation}}
\newcommand{\ee}{\end{equation}}
\newcommand{\bea}{\begin{eqnarray}} 
\newcommand{\eea}{\end{eqnarray}}

\newcommand{\Int}{\mathbb{Z}} 
\newcommand{\Nat}{\mathbb{N}}

\newcommand{\Comp}{\mathbb{C}} 
\newcommand{\Set}{\mathbb{S}} 
\newcommand{\Hb}{\mathbb{H}}

\begin{document}

\title{Perfect transfer of multiple excitations in quantum networks}
\author{T. Brougham} 
\affiliation{Department of Physics, FNSPE, Czech Technical University in Prague, B\v rehov\'a 7, 115 19 Praha 1, Star\'e M\v{e}sto, Czech Republic}
\author{G.~M.~Nikolopoulos}
\affiliation{Institute of Electronic Structure and Laser, FORTH, P.O. Box 1527, Heraklion 71110, Crete, Greece}
\author{I. Jex} 
\affiliation{Department of Physics, FNSPE, Czech Technical University in Prague, B\v rehov\'a 7, 115 19 Praha 1, Star\'e M\v{e}sto, Czech Republic}

\date{\today}

\begin{abstract} 
We present a general formalism to the problem of perfect state-transfer (PST), 
where the state involves multiple excitations of the quantum network. 
A key feature of our formalism is that it allows for inclusion of nontrivial 
interactions between the excitations. Hence, it is perfectly suited to addressing 
the problem of PST in the context of various types of physical realizations.
The general formalism is also flexible enough to account for situations where multiple 
excitations are "focused" onto the same site.
\end{abstract} 


\maketitle

\section{Introduction} 
The problem of designing a passive network of permanently coupled sites, 
which enables a quantum state to be perfectly transferred from one site to 
another, has attracted considerable interest over the last years 
\cite{review,CDEKL04,morePST,NPL04,PST}. One conceptually simple approach to the 
so called perfect-state-transfer (PST) problem is to look for Hamiltonians 
(solutions) that lead to a unitary transformation that permutes the nodes of 
the network at a well-defined time.  It has been shown that this formalism 
allows one to derive infinitely many different PST Hamiltonians \cite{PST}. 
As with most of the literature on PST, this approach has also been formulated 
within the single excitation sector. 
Typically, the state to be transferred pertains to a single quantum particle 
(e.g., electron, atom, photon, etc) which is fully compatible with 
the ``hardware" of the network (e.g., quantum dots, optical lattices, superconducting qubits, etc).  
In this case, as far as the PST problem is concerned, the particular 
character of the information carrier (boson or fermion) does not play 
any role, and is not reflected in the solutions one may derive. 
The situation, however, becomes substantially different when the state to be  
transferred is encoded on two or more particles. In this case, the bosonic or 
fermionic nature of the information carriers is expected to reveal itself, and together 
with interparticle interactions, will impose additional constraints on the physically acceptable solutions.

Our purpose in this work is to show how the formalism of \cite{PST}, can be 
used for engineering quantum networks that allow for the perfect transfer of 
states encoded on many particles (e.g. entangled states). 
As will be shown, direct use of the single-excitation 
formalism  developed in \cite{PST} can be physically unrealistic and places strong 
restrictions on the types of transport that one can perform.  For this reason 
we will consider a second more general approach, which can incorporate the 
bosonic/fermionic character of the information carriers under consideration, 
as well as the interaction between them.  Furthermore, the second approach 
allows one to design networks that can perform tasks that would be impossible by  
simply adopting the single-excitation formalism of \cite{PST}. For 
the sake of simplicity, the following discussion focuses on two-particle states, 
but our results and conclusions can be generalized to many-particle states.  

\section{General formalism} 
\label{GF}
The quantum communication scenario under consideration pertains to a quantum network, 
consisting of $M$ permanently coupled quantum systems, to be referred to 
hereafter as nodes or sites.   Typically, the network is initially prepared in its 
ground state $\ket{\oslash}$.  At $t=0$, the information to be communicated 
is encoded onto the quantum state of a number of nodes (source nodes).
In the context of the PST problem one looks for solutions (PST Hamiltonians)
that guarantee the transfer of the input state from the source to some 
other (destination) sites in a deterministic way. 

Depending on the particular implementation under consideration, the input state, 
may require the doping of the source nodes with excess particles that are 
fully compatible with the hardware of the network. Each particle may involve 
various degrees of freedom besides its position on the network 
(e.g., energy, angular momentum, spin, etc), 
and information is encoded on some of them. For instance, in quantum computing 
schemes based on quantum dots, information can be encoded onto the 
spin state of an {\em excess} electron. The quantum dot may, in principle, 
contain more electrons. However, as long as the quantum dot is in its ground state, the net spin will simply correspond to that of the  
excess electron and the entire system behaves as 
a qubit \cite{nch}.  In the case of atoms (or ions), 
the information can be encoded on their internal states. 

In general, therefore, the initial excitations in the 
network may involve various degrees of freedom.  
In the context of PST problems, one usually assumes that 
all the degrees of freedom (apart from the position of the excitations on the network) 
are preserved throughout the evolution of the system on the time scales of interest, 
as well as that the ground state remains frozen 
\cite{review,CDEKL04,morePST,NPL04,PST,remark1}. 
The problem of PST then boils down to the design of 
Hamiltonians which ensure the faithful (perfect in the absence of disorder and 
dissipation) transfer of all the initial excitations, from the source to the 
destination sites. As long as the Hamiltonians of interest preserve 
the total number $(N)$ of excitations in the system, one can also focus 
on the $N$-excitation Hilbert subspace $\Hb_{N}$, where an orthonormal basis is 
formed by all the states 
$|\{n_{1,\sigma}\};\{n_{2,\sigma}\}; ...;\{n_{M,\sigma}\}\rangle$, 
with $\sum_{j,\sigma}{n_{j,\sigma}}=N$. 
Here, $|n_{j,\sigma}\rangle$ denotes the number of excitations in the state $\sigma$ at 
the $j$th site, where $\sigma$ accounts for all degrees of freedom including the ones 
used for information encoding, and let $\Set$ denote a set of orthonormal basis states.

The case of single excitation (i.e., $N=1$), has been addressed by many authors, 
and various solutions have been obtained \cite{review}. In this context, we have also 
presented a rather general approach to the problem of PST, where solutions are associated with 
a permutation $\mathscr{\hat{P}}$, which permutes all the nodes of the network 
i.e., $\mathscr{\hat{P}}:~j\to \mathscr{\hat{P}}(j)$ \cite{PST}.
The energies and the coupling constants in the Hamiltonian $\mathscr{\hat{H}}$, 
are designed such that the corresponding evolution operator 
$\mathscr{\hat{U}}(t)\equiv e^{-i\mathscr{\hat{H}}t}$, satisfies
\be
\mathscr{\hat U}(\tau)=\mathscr{\hat{P}}.
\label{pst_cond}
\ee
Thus a single excitation that is prepared on the $j$th site at $t=0$, 
will be transferred to the site $\mathscr{\hat{P}}(j)$ at time $\tau$. 
A brief summary of our approach is provided in the appendix.

In the language of second quantization, the initial state of the 
network is of the form 
\be
\ket{\Psi(0)}= f(\{\hat a^{\dagger}_{j,\sigma}\})\ket{\oslash},
\ee
where $f(\{\hat a^{\dagger}_{j,\sigma}\})$ is a function of 
creation operators $\hat a^{\dagger}_{j,\sigma}$, which create excitations 
in state $\sigma$ at the $j$th site. 
The action of a PST Hamiltonian can thus be described in terms of 
permuting the creation operators, so that at time $\tau$ we have
\be
\ket{\Psi(\tau)}= f(\{\hat a^{\dagger}_{\mathscr{\hat{P}}(j),\sigma}\})\ket{\oslash}.
\ee
Given the generality of $f(\{\hat a^{\dagger}_{j,\sigma}\})$, one may argue 
here that a PST Hamiltonian that has been obtained within the single-excitation 
subspace, can be also applied for the transfer of multiple excitations.  
Unfortunately, such an approach, does not take into account possible interactions between 
the particles, and is limited thus only to the description of non-interacting excitations. 
Another limitation of the single-particle formalism is that it cannot describe situations where
two or more excitations, from different nodes, are transferred onto the same node.
This is due to the fact that by definition a permutation will map  
each node of the network to a unique node, i.e. $\mathscr{\hat{P}}(j)=\mathscr{\hat{P}}(k)$,  
if and only if $j=k$. 

It is essential therefore to reformulate our approach to the problem of PST, so that it is also 
applicable to more general scenarios, where multiple excitations are involved. 
Given that the system is restricted to the $N$-excitation Hilbert space 
$\Hb_N$ throughout its evolution, its total wavefunction at any time, can be expressed in 
terms of the basis states $|\{n_{1,\sigma}\};\{n_{2,\sigma}\}; ...;\{n_{M,\sigma}\}\rangle$.  
The key idea is that, for the transfer of $N$ excitations the permutation 
should be defined with respect to these basis states, and 
not with respect to the sites of the network. 
The rank of a permutation that permutes all the nodes is $M$, whereas 
the rank of a permutation that permutes all the basis states is $M^N$. 
It is only in the case of single excitation that the two descriptions coincide. 
As soon as we have a permutation with respect to the  basis states 
$|\{n_{1,\sigma}\};\{n_{2,\sigma}\}; ...;\{n_{M,\sigma}\}\rangle$, one can view 
the problem of PST as if we have a single excitation in a network composed of 
$M^{N}$ sites (each site corresponds to a basis state). 
We can thus use directly the formalism that has been developed in \cite{PST}, 
for the design of PST Hamiltonians within the single-excitation space (see appendix). 

For the sake of clarity, from now on we focus only on the case of two excitations; 
this will considerably simplify our notation. Our main observations and conclusions 
will, however, also be valid for the case of $N>2$.
The two-excitation Hilbert space $\Hb_2$ is spanned by the states 
$\{\ket{{i,\mu};{j,\nu}}\}$, with $\mu,\nu\in\Set$,  
where we have set $\ket{{i,\mu};{j,\nu}}:=\ket{1_{i,\mu},1_{j,\nu}}$. 
Let us now assume that the input state is $\ket{\psi(0)}=\ket{s_1,\mu;s_2,\nu}$, 
which automatically specifies a particular ordering of the states $\{\mu,\nu\}$, 
with respect to their position in the network. We can thus define three different 
subspaces of $\Hb_2$, with respect to this initial ordering, that is 
\begin{subequations}
\label{subspaces}
\bea
\Hb_2^{(<)},~\textrm{spanned by}~\{\ket{i,\mu;j,\nu}~:~i<j\},\\
\Hb_2^{(=)},~\textrm{spanned by}~\{\ket{i,\mu;j,\nu}~:~i=j\},\\
\Hb_2^{(>)},~\textrm{spanned by}~\{\ket{i,\mu;j,\nu}~:~i>j\}.
\eea
\end{subequations}

\begin{figure}
\includegraphics[width=8.cm]{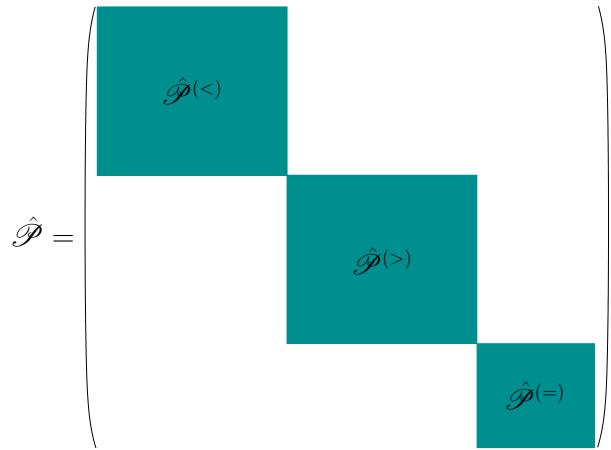}
\caption{(Color online) A permutation of block diagonal form.}
\label{fig1}
\end{figure}

When these subspaces are decoupled, the system is restricted within the subspace 
that is specified by $\ket{\psi(0)}$. 
Hence, the problem of PST is simplified further, as it can be solved 
separately within each subspace. In particular, 
one can consider permutations of the block-diagonal form depicted in Fig. \ref{fig1}. 
The block $\mathscr{\hat{P}}^{(\cdot)}$ is a permutation that permutes all the 
basis states of the corresponding subspace $\Hb_2^{(\cdot)}$. For networks that are 
insensitive to different degrees of freedom $\mu\in\Set$, 
the subspaces $\Hb_2^{(<)}$ and $\Hb_2^{(>)}$ are basically 
equivalent and thus $\mathscr{\hat{P}}^{(<)}=\mathscr{\hat{P}}^{(>)}$.
Starting from the eigenvalues of $\mathscr{\hat{P}}^{(\cdot)}$, 
one can work backwards along the lines of the appendix, 
to design Hamiltonians that generate this permutation at 
time $\tau$, i.e., they satisfy Eq. (\ref{pst_cond}) with 
$\mathscr{\hat{P}}=\mathscr{\hat{P}}^{(\cdot)}$. In the most general scenarios, 
however, the subspaces are not decoupled and cannot be treated separately. 
In these cases, one may also have transfer between states that belong to different 
subspaces, and the PST problem is more involved.

Using the above methodology, one can design PST Hamiltonians in the case of two or more 
interacting as well as non-interacting excitations, which can be either bosons or fermions. 
In the following section we will demonstrate, using some simple examples, how various constraints regarding the nature of the excitations can be incorporated in our formalism. 

\section{Examples}
In the case of noninteracting excitations, the basis state $\ket{i,\mu;j,\nu}$ 
is characterized by the energy,  
\[E_{i,j}^{(\mu,\nu)}=\varepsilon_{i,\mu}+\varepsilon_{j,\nu},\] 
where $\varepsilon_{j,\sigma}$ is the energy of an excitation in state $\sigma$ at the $j$th 
site. 
In the case of interacting particles, however, one has 
\[\tilde{E}_{i,j}^{(\mu,\nu)}=E_{i,j}^{(\mu,\nu)}+U_{i,j}^{(\mu,\nu)}.\]
The additional energy $U_{i,j}^{(\mu,\nu)}$, accounts for repulsive ($U_{i,j}^{(\mu,\nu)}>0$) or 
attractive ($U_{i,j}^{(\mu,\nu)}<0$) interactions, whose magnitude may depend on the 
position of the two excitations in the network, as well as other degrees of freedom. 
When the excitations are associated with fermions and occupy the same site, 
the Pauli principle requires that $\mu\neq\nu$, i.e., the excitations   
must pertain to different states. For instance, the ground state of a 
two-electron quantum dot in zero magnetic field is the spin singlet state, 
where the two electrons occupy the 
lowest orbital and they have antiparallel spins \cite{newcite}.  
By contrast, in the case of bosons, there are no 
such restrictions, and both excitations can be in the same state, while occupying the 
same site. This is, for instance, the case of the so-called dimers, trimers, etc, 
in optical lattices \cite{ca1,ca2,ca3}. Besides onsite interaction, one may also have inter-site 
interactions when the two interacting excitations occupy different sites, but their 
spatial separation is sufficiently small for interactions to occur. Certain 
implementations offer unprecedented controllability on both of these types of 
interactions, allowing thus for various scenarios. 

In the following we consider particular cases, restricting ourselves to 
networks, that are insensitive to different degrees of freedom $\mu\in\Set$. 
Their evolution is governed by time-independent Hamiltonians, that 
preserve the total number of excitations, as well as the initial choice 
of $(\mu,\nu)$ on the time scales of interest, and are of the form 
\bea
\label{generalHam}
\mathscr{\hat{H}}&=&\sum_{i,\sigma}\varepsilon_{i,\sigma}\hat{n}_{i,\sigma}
+\frac{1}{2}\sum_{i,k}\sum_{\sigma,\sigma^\prime}U_{i,k}^{(\sigma,\sigma^\prime)}\hat{n}_{i,\sigma}\big(\hat{n}_{k,\sigma^\prime}-\delta_{i,k}\delta_{\sigma\sigma'}\big)\nonumber\\
&+&\sum_{i<k}\sum_\sigma J_{i,k}(\hat{a}_{i,\sigma}^\dag\hat{a}_{k,\sigma}+h.c),
\eea
where $\hat{n}_{i,\sigma}=\hat{a}_{i,\sigma}^\dag\hat{a}_{i,\sigma}$. 
The last term of 
the Hamiltonian describes the hopping of the excitation between various 
sites in the network, with the corresponding set of coupling constants 
$\{J_{i,k}\}$, taken to be insensitive to $\sigma$. 
Such types of Hamiltonians are met in various contexts.

\subsection{Decoupled subspaces}
In various realistic scenarios, the onsite interaction $U_{j,j}^{(\mu,\nu)}$ 
is much larger (in absolute value) than any other energy 
scale in the system \cite{footnotenew}.  
In such cases, the energy gap between the basis states $\{\ket{i,\mu;j,\nu}\}$ 
and $\{\ket{j,\mu;j,\nu}\}$ is of the order of $U_{j,j}^{(\mu,\nu)}$. 
Assuming further non-vanishing couplings only between nearest neighbors (NN) in Eq. (\ref{generalHam}) 
(i.e., $J_{i,k}\neq 0$ for $i=k\pm 1$), transitions between states 
$\{\ket{i,\mu;j,\nu}\}$ with $i<j$ and $i>j$ are only possible 
via the states $\{\ket{j,\mu;j,\nu}\}$ which are far off-resonant. 
Hence, if the time scale over which PST has to be achieved satisfies 
\[\tau\ll U_{j,j}^{(\mu,\nu)}/J_{j-1,j}^2,\] 
the subspaces of Eq. (\ref{subspaces}) can be considered, to a good approximation, decoupled 
throughout the evolution of the system. Various examples will now be investigated.

Suppose that the system is initially prepared in the state  
$\ket{\psi(0)}=\ket{s_1,\mu; s_2, \nu}$, for $\mu,\nu\in\Set$ and $s_1<s_2$. 
Given that the system is restricted within the subspace $\Hb_2^{(<)}$ 
throughout its evolution, the initial ordering of 
$\{\mu,\nu\}$ will be preserved, whereas the states of the other 
subspaces, are practically forbidden.
We will thus focus on the subspace $\Hb_2^{(<)}$. The corresponding sub-permutation, $\mathscr{\hat{P}}^{(<)}$, will be 
determined by the definition of the destination nodes $(d_1,d_2)$ as well as any additional 
physical restrictions imposed on the system. 

For instance, if one is interested in PST Hamiltonians with mirror symmetry, 
the permutation $\mathscr{\hat{P}}^{(<)}$ has to have the antidiagonal form 
\be
\label{perm}
\mathscr{\hat{P}}^{(<)}=\sum_{i=1}^M\sum_{j>i}\ket{i,\mu;j,\nu}\bra{(M+1-j),\mu;(M+1-i),\nu},
\ee 
where $i,j\in\{1,2,...,M\}$.  Working along the lines of the appendix, 
one may find parameters $\{\varepsilon_{i,\sigma}, U_{i,k}^{(\sigma,\sigma^\prime)}, J_{i,k}\}$ 
such that the Hamiltonian (\ref{generalHam}) satisfies Eq. (\ref{pst_cond}), 
within the subspace $\Hb_2^{(<)}$. 

In order for the Hamiltonian (\ref{generalHam}) to be compatible with the 
permutation (\ref{perm}), one has to ask for 
\be
\label{com}
[\mathscr{\hat{P}}^{(<)}, \mathscr{\hat{H}}]=0,
\ee 
within the subspace of interest spanned by the states $\{\ket{i,\mu;j,\nu}~|~i<j\}$.
This requirement imposes the first constraints on the choices of 
$\varepsilon_{i,\sigma}$, $J_{i,k}$, and $U_{i,k}^{(\sigma,\sigma^\prime)}$ namely, 
\begin{subequations}
\label{cond1}
\bea
&&J_{i,k}=J_{M+1-k,M+1-i},\\
&&\tilde{E}_{i,k}^{(\mu,\nu)}=\tilde{E}_{M+1-k,M+1-i}^{(\mu,\nu)}.
\eea
\end{subequations}
Additional constraints, depend on the details of the network one is interested in. 
In the absence of any intersite repulsive (or attractive) interactions    
(i.e.,  $U_{i,k}^{(\mu,\nu)}=0$ $\forall$ $i,k\in\{1,2,...,M\}$ and $i\ne k$), 
one of the solutions pertains to \cite{remark2}
\bea
\varepsilon_{i,\mu}=\varepsilon_{k,\nu},\quad\textrm{and}\quad
J_{i,i+1}\propto\sqrt{i(M-i)},
\eea
This is not a new result in the context of two excitations, as it has been predicted by one of us 
in an earlier collaboration \cite{NPL04}, and its robustness against various types of imperfections 
has been analyzed in \cite{NPL04,PNPL10}. We should emphasize, however, that this is only one of the infinitely many solutions 
that one may obtain within the present theoretical framework, for the particular choice 
of $\mathscr{\hat{P}}^{(<)}$. 

\begin{figure}
\center{\includegraphics[width=9cm,height=!]
{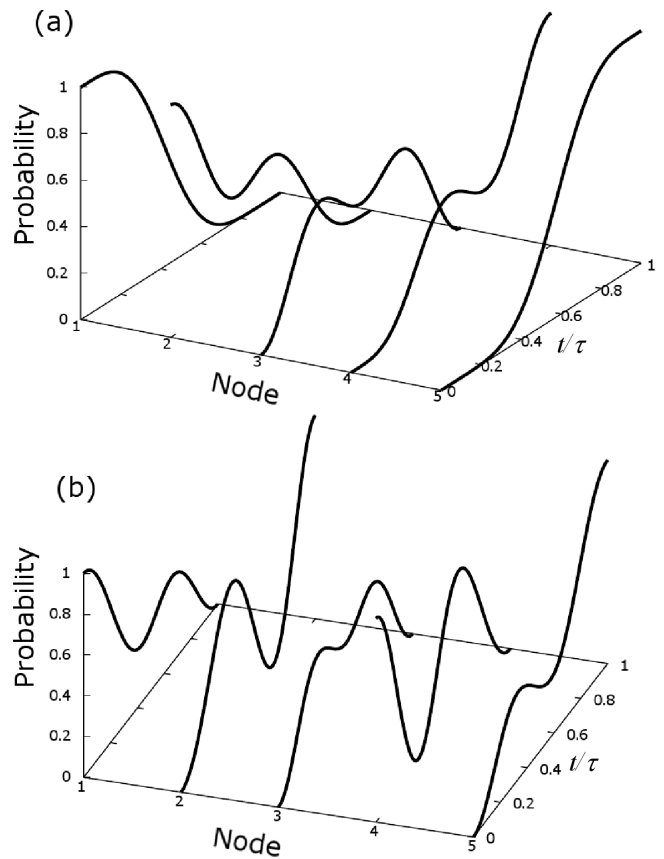}
\caption{Plots of the probability of finding an excitation at each node as a function of time, for a network of $M=5$ sites.  
Both plots correspond to the parameters $W=0.035672$, $J_{1,2}=3.15891$ and $J_{2,3}=3.84084$.  
In plot (a) the initial state $|1,\mu;2,\nu\rangle$ is transferred to $|4,\mu;5,\nu\rangle$, while in (b) 
the initial state $|1,\mu;4,\nu\rangle$ is transferred to $|2,\mu;5,\nu\rangle$.}
\label{fig2}}
\end{figure}

Relaxing the constraint $U_{i,k}^{(\mu,\nu)}= 0$ for $i\neq k$, 
allows one to take account of repulsive(attractive) interactions between excitations 
e.g., in adjacent sites only (i.e. $U_{i,k}^{(\mu,\nu)}\ne 0$, for $k=i\pm 1$).  
Let us assume that this interaction does not change with $i$, hence we set $U_{i,i+1}^{(\mu,\nu)}=W$.
In this case, the derivation of 
analytic expressions is far from trivial, but one can always resort to numerical solutions. 
Given the centrosymmetry as well as the assumption of NN couplings, 
the parameters to be estimated are 
$\{\varepsilon_{i,\mu},\varepsilon_{k,\nu},J_{1,2},J_{2,3},W\}$.  
In table \ref{tab:table1}, we provide a number of solutions for the case of a 
network consisting of $M=5$ sites, with 
 $\varepsilon_{i,\mu}=\varepsilon_{k,\nu}$, 
$\forall i,k\in\{1,\ldots,M\}$.  A sample of the dynamics of the transfer is given in figure \ref{fig2}.

\begin{table}
\caption{\label{tab:table1}
Parameters for perfect transfer of two excitation in a network consisting of $M=5$ sites,  
with mirror symmetry and NN interactions. Other parameters: $\varepsilon_{i,\mu}=\varepsilon_{k,\nu}$, 
$J_{1,2}=J_{4,5}$, and $J_{2,3}=J_{3,4}$.  The results were obtained for a Hamiltonian with prescribed eigenvalues $\epsilon^{(a)}_{\lambda_n}/\tau=\pi\times\{\pm 3, \pm2, \pm 1,\pm 1, 0, 0\}$ (see appendix).}
\begin{ruledtabular}
\begin{tabular}{ccc}
$W$ & $J_{1,2}$ & $J_{2,3}$\\ 
\hline
0.035672 & 3.15891 & 3.84084\\
0.0584814 & 3.13976 & 3.84777\\ 
0.0651748 & 3.15557 & 3.84182\\ 
0.0765194 & 3.14260 & 3.84542\\  
0.131893 & 3.0849 & 3.87000\\ 
\end{tabular}
\end{ruledtabular}
\end{table}

For all of the solutions in table \ref{tab:table1}, $W<J_{i,k}$.  When $W$ is greater than the coupling constants we cannot obtain 
solutions with NN couplings, unless it is possible for two excitations to occupy a single site; something we do 
not allow in this section.  The reason for this is that for NN couplings the states $|m,\mu;m+1,\nu\rangle$ 
and $|m+1,\mu;m+2,\nu\rangle$ are not coupled directly, but only by via the state $|m+1,\mu;m+1,\nu\rangle$.  
When $W>J_{i,k}$ transitions such as $|m,\mu;m+1,\nu\rangle$ to $|m,\mu;m+2,\nu\rangle$ become off resonant. 

As long as we are interested in networks that are insensitive to different 
degrees of freedom, the subspaces $\Hb_2^{(<)}$ and $\Hb_2^{(>)}$ are  
equivalent, and the case of $s_1>s_2$ is covered by the previous one.

\subsection{Transitions between subspaces}
Our examples up to now were for PST Hamiltonians with NN couplings only. 
There are, however, realistic situations where the effects of couplings beyond NN 
cannot be ignored.  Formally speaking, the presence of non-vanishing couplings 
beyond NN in Eq. (\ref{generalHam}), automatically enables transitions 
between different subspaces, and thus the permutations one may consider cannot 
be expressed in the block diagonal form of Fig. \ref{fig1}.

\begin{figure}
\includegraphics[width=8.cm]{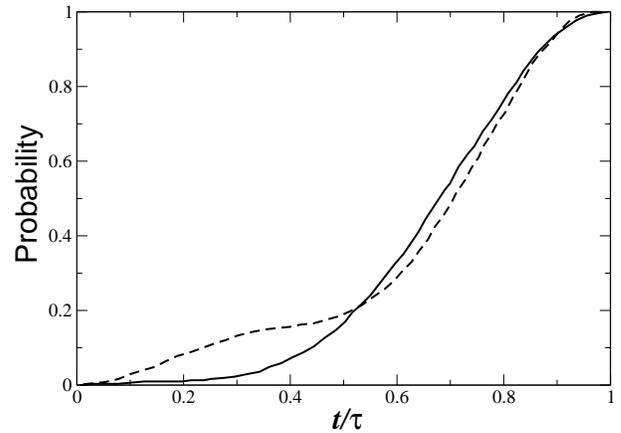}
\caption{
Plot of the probability of transfer as a function of time, for a centrosymmetric network with $M=5$ sites, and spatially-dependent couplings of the form 
(\ref{nonNNcouple}), with $J=\pi$. The parameters used are those given in table \ref{tab:table2}.  The solid line shows the probability of finding the system in the state 
$|4,\nu;5,\mu\rangle$ when it was initially prepared in the state $|1,\mu;2,\nu\rangle$.  The dotted line shows the probability of finding the 
system in the state $|3,\nu;4,\mu\rangle$ when it was initially prepared in the state $|2,\mu;3,\nu\rangle$.}
\label{fig3}
\end{figure}

To demonstrate this fact consider, for instance, the transfer of two excitations on 
a centrosymmetric network of $M=5$ sites as described by the transformation
\bea
|i,\mu;j,\nu\rangle\rightarrow |(6-i),\mu;(6-j),\nu\rangle.
\label{trans}
\eea 
Note here that for $i\lessgtr j$ the system is transferred to the subspace $\Hb_2^{(\gtrless)}$.
Assume that the network has a linear geometry with spatially dependent interactions of the form  
\begin{eqnarray}
\label{nonNNcouple}
&&J_{1,2}=\frac{J}{r_1},\,\,J_{1,3}=\frac{J}{r_1+r_2},\,\,J_{1,4}=\frac{J}{r_1+2r_2},\nonumber\\
&&J_{1,5}=\frac{J}{2r_1+2r_2},\,\,J_{2,3}=\frac{J}{r_2},\,\,J_{2,4}=\frac{J}{2r_2},
\end{eqnarray}
where $r_{1(2)}$ are the distances between the sites 1(3) and 2, while the centrosymmetry  
fixes the remaining couplings in the same manner as in Eqs. (\ref{cond1}). 
As before, double occupancy is neglected (a scenario with double occupancy 
will be discussed later on), while 
we allow for non-vanishing intersite interactions between neighboring sites only, 
i.e. $U_{i,i\pm 1}^{(\mu,\nu)}=W\ne0$ and $U_{i,k}^{(\mu,\nu)}=0$ for $k\ne i+1$ .  
To make the example more interesting, let us suppose that the detunings between the single-particle energies
are such that, the evolution of the system on the time scales of interest is practically restricted 
within the subspace spanned by the two-particle basis states for which  $i+j$ is an odd number i.e., 
$\{|1,\sigma;2,\sigma'\rangle,|1,\sigma;4,\sigma'\rangle,|2,\sigma;3,\sigma'\rangle,|2,\sigma;5,\sigma'\rangle,|3,\sigma;4,\sigma'\rangle,|4,\sigma;5,\sigma'\rangle\}$,
with $\sigma,\sigma^\prime\in\{\mu,\nu\}$. We are interested in the derivation of an effective PST 
Hamiltonian that acts on this subspace and leads to the transformation (\ref{trans}) at time $\tau$.
To this end, our permutation can be defined with respect to the "odd" basis states only, 
neglecting the far off-resonant "even" ones: $|1,\sigma;3,\sigma'\rangle$ , 
$|1,\sigma;5,\sigma'\rangle$, $|2,\sigma;4,\sigma'\rangle$, and $|3,\sigma;5,\sigma'\rangle$. 
The open parameters in the Hamiltonian that need to be determined are 
$\{\varepsilon_{1,\sigma},\varepsilon_{2,\sigma},\varepsilon_{3,\sigma}, J, r_1, r_2, W\}$. 
In table \ref{tab:table2} we present a solution that has been obtained numerically from 
the permutation 
\be
\label{perm2}
\mathscr{\hat{P}}=\sum_{i=1}^5\sum_{\substack{j=1\\ (j\neq i)}}^5\Lambda_{i,j}\ket{(6-i),\mu;(6-j),\nu}\bra{i,\mu;j,\nu},
\ee 
where $\Lambda_{i,j}=1$, for $i+j=2\alpha+1$ with $\alpha\in\Nat$, and $0$ otherwise. 
The corresponding time evolution of the probability of transfer 
for two different initial states is given in Fig. \ref{fig3}. Clearly, in both cases the probability becomes equal to one, and thus transformation (\ref{trans}) is 
achieved, at time $\tau$, although the intermediate dynamics are different. 

\begin{table}
\caption{\label{tab:table2}
Numerically obtained parameters for a centrosymmetric PST Hamiltonian with spatially-dependent couplings  of the form 
(\ref{nonNNcouple}), with $J=\pi$.  The transfer time is taken to be $\tau=1$.}
\begin{tabular}{|| c | c || c | c ||}
\hline
$\varepsilon_{1,\sigma}$ & 7.72530 & $W$ & 1.06485$\times10^{-2}$\\ \hline
$\varepsilon_{2,\sigma}$ & -28.0330 & $r_1$ & 8.91994$\times10^{-1}$\\ \hline
$\varepsilon_{3,\sigma}$ & 6.80085 & $r_2$  & 9.16677$\times10^{-1}$ \\\hline
\end{tabular}
\end{table}

From the perspective of quantum control a useful and important task would be to perfectly transfer two excitations 
that are in different sites onto the same site and vice-versa  (i.e.,  $|i,\mu;j,\nu\rangle\leftrightarrow|k,\mu;k,\nu\rangle$), 
coupling thus the subspaces $\Hb_2^{(\lessgtr)}$ and $\Hb_2^{(=)}$.  This would allow one to perform simple logical operations within 
the quantum network. 
For instance, a triple quantum dot can be used for the generation of spatially separated 
spin-entangled electrons \cite{SarLossPRL03}. Given that the ground state of two excess electrons in a quantum dot 
is the spin-singlet state, with both electrons having the same orbital energy, 
the key issue is to separate spatially the two electrons, by transferring them coherently and jointly, to different adjacent quantum dots. 
As has been shown in \cite{SarLossPRL03}, this can be achieved by operating the network in the Coulomb 
blockade regime. 
We will demonstrate here how one can obtain within our theoretical framework, the effective Hamiltonian 
governing the dynamics of the two electrons in the Coulomb blockade regime, and in the absence of dissipation.

Following the notation introduced in Sec. II, the basis states of interest are $|i,\mu;j,\nu\rangle$, 
with $i,j\in\{1,2,3\}$ and the possible spin states $\mu,\nu\in\{\uparrow, \downarrow\}$. Initially, we 
have two electrons on site 2, with opposite spins i.e., the state of the system is 
$|2,\uparrow;2, \downarrow\rangle$ \cite{remark3}. 
We will assume NN interactions, with the corresponding coupling strengths as well as the onsite repulsions,  
sufficiently small, so that only lowest single-particle states (with energies $\varepsilon_i$) 
are of importance \cite{SarLossPRL03}. The energy corresponding to the state $|i,\mu;j,\nu\rangle$ is then 
$\tilde{E}_{i,j}=\varepsilon_i+\varepsilon_j+U_{i,j}$, with $\varepsilon_{i,\mu}=\varepsilon_{i}$, $U_{i,j}^{(\mu,\nu)}=U_{i,j}$ \cite{remark4}. 
To suppress the single-electron transport in favor of the two-electron transport, one may choose 
$\varepsilon_{i}$ and $U_{i,i}$ so that $|\tilde{E}_{i,i+1}-\tilde{E}_{2,2}|\gg J_{i,i+1}$ and  
$\tilde{E}_{1,3}=\tilde{E}_{22}$, where $J_{i,i+1}$ denotes the coupling strength between the states 
$|2,\uparrow;2, \downarrow\rangle$ and $|i,\uparrow;i\pm 1, \downarrow\rangle$. Moreover, to ensure spatial 
separation of the two electrons, one has to satisfy that $\tilde{E}_{i,i}\neq \tilde{E}_{2,2}$. 
Structures of quantum dots, as well as other realizations, allow for the external adjustment of all the 
relevant parameters, so that these conditions are satisfied simultaneously \cite{newcite,SarLossPRL03}. 
In such case, the off resonant states do not play a 
significant role, and the dynamics of the two electrons are governed by an effective Hamiltonian 
of the form
\begin{widetext} 
\bea 
\label{simplefocus}
\mathscr{\hat{H}}_{\rm eff}&=&\tilde{E}_{22}(\ket{2,\uparrow;2, \downarrow}\bra{2,\uparrow;2, \downarrow}+
\ket{1,\uparrow; 3, \downarrow}\bra{1,\uparrow; 3, \downarrow}
+\ket{1,\downarrow; 3, \uparrow}\bra{1,\downarrow; 3, \uparrow})\nonumber\\
&&+J^{(2)}\big[\ket{2,\uparrow;2, \downarrow} \bra{1,\uparrow; 3, \downarrow} 
+ \ket{2,\uparrow;2, \downarrow} \bra{1,\downarrow; 3, \uparrow} +h.c.\big]. 
\eea
\end{widetext}
Hence, the state $|2,\uparrow;2, \downarrow\rangle$ is directly coupled to the states 
$\{|1,\uparrow; 3, \downarrow\rangle, |1,\downarrow; 3, \uparrow\rangle\}$, 
via the second-order coupling strength $J^{(2)}$.
The present PST formalism is capable of providing us with such a type of Hamiltonians. 
Given, however, that we wish to couple $|2,\uparrow;2, \downarrow\rangle$ to 
both $|1,\uparrow; 3, \downarrow\rangle$ and $|1,\downarrow; 3, \uparrow\rangle$, 
we start with the unitary  
\[\mathcal{P}=\left [\left (|1,\uparrow;3,\downarrow\rangle+|1,\downarrow;3,\uparrow\rangle\right )
\langle 2,\uparrow;2,\downarrow|+h.c.\right ]/\sqrt{2},
\]
rather than a permutation. Using the procedure outlined in the appendix, one obtains a class of Hamiltonians 
of the form (\ref{simplefocus}) with $\tilde{E}_{22}=\pi\Delta_+/(4\tau)$ and $J^{(2)}=\pi\Delta_-/(2\tau)$, 
where $\Delta_{\pm}=2x_+\pm(2x_--1)$ and $x_{\pm}\in\mathbb{N}$.  A Hamiltonian with these parameters will, 
at time $t=\tau$, transform the state $|2,\uparrow;2,\downarrow\rangle$ into the state 
$(|1,\uparrow;3,\downarrow\rangle+|1,\downarrow;3,\uparrow\rangle)/\sqrt{2}$, producing thus a pair 
of spatially separated spin-entangled electrons. For any choice of the integers $x_{\pm}$ , we  will obtain  
an effective Hamiltonian that performs the desired state transfer, apart from an unimportant global phase.

The previous scenario can be placed in the context of different physical realizations, which instead of fermions 
may involve bosonic excitations prepared in different internal states  \cite{remark5}.  
In the case of bosons, however, one may also consider the case where the two excitations are in the same state 
$\mu$ 
(e.g., photons with the same polarization, or atoms in the same internal state), and look for Hamiltonians that perform 
the  transformation $|1,\mu;3,\mu\rangle\to|2,\mu;2,\mu\rangle$, and vice versa.  
To describe this scenario in our formalism, we take the permutation to be 
\[\mathcal{P}=|1,\mu;3,\mu\rangle\langle 2,\mu;2,\mu|+h.c.,\]
and using the procedure outlined in the appendix, we obtain again a Hamiltonian of the form (\ref{simplefocus}), 
with $\tilde{E}_{22}=\pi\Delta_+/(4\tau)$ and  $J^{(2)}=\pi\Delta_-/(2\tau)$.

In closing, we would like to point out that in the last two examples our formalism was used for the derivation 
of an effective Hamiltonian which acts on a subspace of near-resonant states. 
This effective Hamiltonian is basically an approximation of the original Hamiltonian, and the state transfer is thus not perfect, 
but rather occurs with high fidelity.
How large the fidelity can be made depends on to what extent the system remains in the particular subspace throughout its 
evolution on the time scale of interest. Ideally, for a network whose components allow for arbitrary adjustment of all the related 
parameters (such as couplings, detunings, interactions, etc), the actual evolution of the system can be brought arbitrarily close 
to the effective one, and thus the fidelity can be arbitrarily close to 1 (in the absence of disorder and dissipation).      
In this direction, quantum dots \cite{newcite} have been shown to offer unprecedented controllability 
in many respects, although 
deviations from the ideal case are to be expected. Further discussion on these issues is beyond the scope of this work, since it requires 
a thorough consideration of the various components of the network. 

\section{Conclusions}
We have extended the PST formalism of \cite{PST}, to states that are encoded on multiple excitations within a quantum network. This allows us to describe situations such as transferring a singlet state encoded on two excess electrons that are in neighboring quantum dots, within in an extended array of quantum dots. The starting point of the 
present generalized approach is a permutation (or, in general, a unitary transformation) which is defined with respect to the many-body basis states, and permutes the initial and the target states.  Working backwards, one can construct infinitely many different PST Hamiltonians that generate the particular permutation at a well defined time. We have also demonstrated how our formalism can be used for the derivation of an effective Hamiltonian. Given a particular physical setup, one has first to identify the far off-resonant (many-body) states which, as a result of various constraints, are scarcely populated throughout the evolution of the system. The above mentioned transformation should be defined within the subspace spanned by the remaining states that are near resonant with the initially populated one.  
In this case, the rank of the transformation is smaller than $M^N$, and it is determined by the dimension of the subspace spanned by the
states of interest. Clearly, one cannot make any definite statements about the value of the rank or how it scales 
with the number of particles, since they both depend on the particular physical realization under consideration, and the time scale over
which the transfer has to take place (determined by the coupling constants between different sites). For instance, according to
Pauli's principle two fermions can never occupy the same state, and such states can be excluded immediately from our analysis.
On the other hand, two bosons may be prevented from occupying the same site on a certain time scale, when the repulsion between them is sufficiently large. 
There are cases, therefore where certain states can be excluded automatically from our analysis due to fundamental physical principles, and cases 
where all of the states are, in principle, allowed and which ones can be excluded (as an approximation) depends entirely on the details of the 
network and the time scales of interest.
  
Although our methodology, can be applied to quantum networks with any number of excitations throughout this work,  for the sake of illustration and simplicity,  we have focused on two excitations only.  We have demonstrated how one can take into account constraints pertaining to the particular physical realization under consideration, as well as the nature of the excitations and the interactions among themselves. The examples we have discussed are by no means exhaustive, but they do provide the guidelines for imposing physical constraints in our approach.

\section*{Acknowledgements}
T.B. and I.J. acknowledge financial support from the Doppler Institute and from grants MSM6840770039 and MSMT LC06002 of the Czech Republic.  G. M. N. acknowledges support from the EC RTN EMALI (contract No. MRTN-CT-2006-035369).

\begin{appendix} 
\section{Summary of the method}
Consider the problem of PST in a Hilbert subspace 
spanned by the states $\{\ket{\psi_j}\}$, with 
$0\leq j\leq L-1$. We are interested in Hamiltonians 
that take the input state $\ket{\psi_s}$ 
to the output state $\ket{\psi_d}$ at time $\tau$. 
We start from a permutation that permutes 
the two states i.e., 
\be
\mathscr{\hat{P}} 
= \ket{\psi_d}\bra{\psi_s}+\mathscr{\hat{P}}^\prime, 
\label{eq:generalP} 
\ee 
where $\mathscr{\hat{P}}^\prime$ is also a  permutation in the basis $\{\ket{\psi_j}\}$.  

In general, $\mathscr{\hat{P}}$ can be decomposed into disjoint cycles $\hat{\mathscr{P}}_i$, 
i.e., $\mathscr{\hat{P}}=\sum_i\hat{\mathscr{P}}_i$. Let 
$\Omega_i$ denote the set of basis states permuted by $\hat{\mathscr{P}}_i$. 
The sets $\{\Omega_i\}$ are disjoint, while the dimensions of the support of each cycle 
$\hat{\mathscr{P}}_i$, denoted by $L_i$, will be the cardinality of the corresponding set $\Omega_i$. 

The spectrum of each cycle $\hat{\mathscr{P}}_i$ is nondegenerate and let $|v_{i}^{(\lambda_n)}\rangle$ 
be the eigenvector corresponding to the eigenvalue $\lambda_n$, 
where $\hat{\mathscr{P}}_i|v_{i}^{(\lambda_n)}\rangle=\lambda_{n}|v_{i}^{(\lambda_n)}\rangle$, with 
\bea
\ket{v_{i}^{(\lambda_n)}}=\frac{1}{\sqrt{L_i}}\sum_{\kappa\in\Omega_i}\lambda_n^{\zeta_\kappa}\ket{\kappa},
\label{eq:McycleEVector} \eea  
and
\bea \lambda_n=\exp\left ({\rm i}2\pi\frac{n}{L_i}\right )\,\textrm{for}\, n\in\Int_{L_i}\equiv\{0,1,\ldots, L_i-1\}.
\label{eq:McycleSpectrum}\eea 
The elements of the set $\Omega_i$ are considered to be arranged in ascending order, 
and $\zeta_\kappa\in \Int_{L_i}$ is the position of the element $\kappa\in\Omega_i$. 
Hence, for the permutation $\hat{\mathscr{P}}$, the eigenvalue $\lambda_n$ corresponds to 
$\eta_{\lambda_n}$ distinct eigenvectors $\{|v_{i}^{(\lambda_n)}\rangle\}$, 
with $i$ running only on the various cycles having $\lambda_n$ in common.

A class of PST Hamiltonians that satisfy Eq. (\ref{pst_cond}) is of the form   
\bea \hat{\mathscr{H}}_{\mathbf{x}}=\frac1\tau\sum_{\lambda_n}
\sum_{a=1}^{\eta_{\lambda_n}}\epsilon_{\lambda_n}^{(a)}
\ket{y_{\lambda_n}^{(a)}}\bra{y_{\lambda_n}^{(a)}},
\label{deg_Ham}
\eea 
where $\epsilon_{\lambda_n}^{(a)}=-\arg\left (\lambda_n\right )+ 2\pi x_{\lambda_n}^{(a)}$ and 
$\mathbf{x}\in\Int^L\equiv\{(x_{\lambda_0}^{(1)},\ldots,x_{\lambda_0}^{(\eta_{\lambda_0})};x_{\lambda_1}^{(1)},\ldots,x_{\lambda_1}^{(\eta_{\lambda_1})};\ldots)~|~x_{\lambda_n}^{(a)}\in\Int\}$. 
For a given eigenvalue $\lambda_n$, the $\eta_{\lambda_n}$ distinct vectors 
$\{|y_{\lambda_n}^{(a)}\rangle\}$ form an orthonormal basis for the 
corresponding subspace and are of the form 
\bea
\ket{y_{\lambda_n}^{(a)}}=\sum_{i} \beta_{a,i}^{(\lambda_n)}\ket{v_{i}^{(\lambda_n)}},
\label{y_na}
\eea
with $\beta_{a,i}^{(\lambda_n)}\in\Comp$ and $\sum_i\beta^{(\lambda_n)*}_{a,i}\beta_{a^\prime,i}^{(\lambda_n)}=\delta_{a,a^\prime}$.  
\end{appendix}


\begin{thebibliography}{00}
\bibitem{remark1} When comes to the realization 
of a PST Hamiltonian in a given system, one has to ensure that these assumptions are 
fulfilled.
\bibitem{footnotenew} In the same fashion (i.e., by setting $U_{j,j}^{(\mu,\mu)}\rightarrow \infty$) one can formally describe the Pauli principle, which does not allow two electrons with the same state, to occupy the same site.
\bibitem{remark2} The actual value of the energies does not play any role in this solution, 
as it will give rise to an unimportant global phase.
\bibitem{remark3} Recall here that, $|i,\mu;j,\nu\rangle$ 
is an abbreviation for the number state $\ket{1_{i,\mu};1_{j,\nu}}$, which does not provide any information 
about the particular 
spin state of the two electrons, apart from the fact that the two electrons have opposite spins. 
For the reasons discussed earlier, the actual spin state of the two electrons is 
$(\ket{\uparrow_2, \downarrow_2}-\ket{\downarrow_2,\uparrow_2})/\sqrt{2}$.
\bibitem{remark5} In this case, the total wavefunction has to be symmetric and not antisymmetric.
\end{thebibliography}
\end{document}